# Amorphous silicon-based microchannel plate detectors with high multiplication gain

J. Löffler, C. Ballif, N. Wyrsch

École Polytechnique Fédérale de Lausanne (EPFL), Institute of Microengineering (IMT), Photovoltaics and Thin Film Electronics Laboratory, Rue de la Maladière 71b, CH-2002 Neuchâtel, Switzerland

Abstract

With their fast response time and a spatial resolution in the range of a few microns, microchannel plates (MCPs) are a prominent choice for the development of detectors with highest resolution standards. Amorphous silicon-based microchannel plates (AMCPs) aim at overcoming the fabrication drawbacks of conventional MCPs and the long dead time of their individual channels. AMCPs are fabricated via plasma deposition and dry reactive ion etching. Using a state-of-the-art dry reactive ion etching process, the aspect ratio, so far limited to a value of 14, could be considerably enhanced with a potential for very high gain values. We show first fabricated AMCP devices and provide an outlook for gain values to be expected based on the fabrication results.

## 1. Introduction

The performance of microchannel plates (MCPs) [1] is defined by inherent properties of their base material, which is typically lead glass. MCPs should be conductive enough to replenish charges in the channel walls while sustaining a high electric field. Other than the channel geometry, the secondary emission of the channel wall largely determines the level of electron multiplication gain.

In order to simplify their cumbersome fabrication process, the trend for novel MCPs is to become independent of their bulk material, since channel walls can be functionalized with coatings of high secondary emissive layers deposited by atomic layer deposition (ALD) [2]. The bulk material, however, can play a key role in order to overcome some critical limitations of MCPs regarding their transient and dead time behavior and to offer alternative MCP fabrication possibilities.

Amorphous silicon-based MCPs (AMCPs) have recently been proposed as microchannel plates using hydrogenated amorphous Si (a-Si:H) as the bulk material [3]. The conductivity of a-Si:H of about $10^{-11}$ $\Omega^{-1}$cm$^{-1}$ can easily be changed via doping by 10 orders of magnitude to tune the conductivity of the AMCP. a-Si:H is a radiation-hard material, which could be beneficial in terms of lifetime and sensitivity to a wide range of irradiation flux. AMCPs can be monolithically integrated on the readout chip as the a-Si:H layer stack is deposited. AMCPs can be fabricated using modern silicon technologies, where the resolution is constantly improving and the area is being scaled up [4].

AMCPs are deposited by plasma-enhanced chemical vapor deposition (PECVD) and structured using a lithography process for channel etching. Channels are micro-machined via dry reactive ion etching (DRIE). So far, their aspect ratio (channel length over channel diameter) has been limited to about 14 [5]. In this paper, we show that the use of a state-of-the-art DRIE process this aspect ratio value can increase considerably allowing for high multiplication gains.

## 2. AMCP architecture

A MCP is an array of pores where both ends are coated with an electrode in order to apply a high electric field across the channels. In a detector assembly, the MCP is mounted above the anode pads. A vacuum space separates the MCP and the anode. In contrast, AMCPs are monolithically deposited on the anode pads that are connected to the readout electronics. The AMCP stack comprises of a stack of active micro channels with electrodes on both sides, where the avalanche is created, and an insulating part between the multiplication region and the anodes.

The architecture of an AMCP is shown in figure 1a. Cr pads deposited on an oxidized Si wafer are collecting the electron avalanche signals from the channels. A highly resistive layer of 2 µm a-Si:H is deposited on the bottom electrodes isolating the anodes from the channel electrodes. This decoupling layer allows for minimization of dark current on the readout pads. A conductive 100 nm layer of n-doped µc-Si is deposited on top of the decoupling layer as one of the channel electrodes. This intermediate electrode between the multiplication part and the decoupling part evacuates the leakage current created by the high electric field. Figure 1c shows an SEM cross section of the layer stack up to the intermediate electrode with an adhesion layer above it.

A thick layer of a-Si:H (of about 80 µm in the present case) forms the multiplication part of the AMCP. a-Si:H stacks with thicknesses in excess of 100 µm have been deposited so far. On top of the stack, an n-doped µc-Si:H layer is used as a top electrode, which in the latest device is replaced by the Cr layer used as a hard mask for the DRIE. Micro channels are then etched through the whole stack, down to the anode pads. Figure 1b shows an SEM image of the top of the AMCP channels. So far, channel diameters have been limited to about 6 µm.

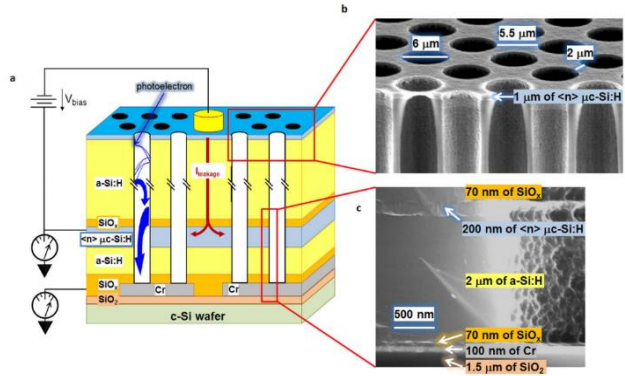

*Figure 1 : a) Schematic representation of the AMCP architecture. Cr pads on an oxidized Si wafer act as the anode to collect the electron avalanches. 2 µm a-Si:H deposited on a $SiO_x$ adhesion layer separate the multiplication layer and signal collection. About 80 µm of a-Si:H deposited on a second SiOx adhesion layer act as the multiplication layer. An n-doped µc-Si:H layer, presently replaced by a Cr layer is used as top electrode. b) SEM image of micro channels and their critical diameters in state-of-the-art AMCPs. c) SEM cross section of the employed layer structure. From* [5].

3. Fabrication of AMCP with high aspect-ratio

In order to use the newest generation of DRIE, AMCPs presented here are fabricated on 6″ wafers. Thus, the fabrication area has been enlarged compared to the formerly fabricated AMCPs on 4″ wafers. A schematic of the fabrication process is shown in Figure 2.

For the bottom electrode, an oxidized Si wafer is coated with a 100 nm layer of Cr (yellow) via sputtering, see step 1. The wafer is dried for 30 min at 200°C, then coated with 1.5 µm photoresist AZ1518 and soft-baked at 85°C for 1 min. It is exposed using a mask aligner, developed in AZ351B 1:4 H2O and post-baked at 85°C for 30 min. The Cr structure is etched with a solution of Ceric ammonium nitrate : perchloric acid : $H_2O$ in a ratio of 10.9 % : 4.25 % : 84.85 % for 250 s. The resist is stripped before the deposition of subsequent layers via PECVD. The patterned wafer is shown in step 2.

During plasma depositions, the wafer is kept at 205°C, unless stated otherwise. The wafer is first cleaned with a $H_2$ plasma for 5 min. After a 20 nm adhesion layer of $SiO_x$ (beige), the 2 µm a-Si:H (light blue) decoupling layer is deposited. n-doped microcrystalline silicon (µc-Si, dark blue) as the intermediate electrode is deposited on top of the stack. After contacting the intermediate electrode locally at the periphery with a 80 nm thick Cr pad, an adhesion layer of 80 nm $SiO_x$ is deposited, followed by the thick a-Si:H multiplication layer. For an optimized quality of this layer, the temperature is changed gradually to 210°C within the first 7 h and kept constant for the remaining 5 h of deposition time.

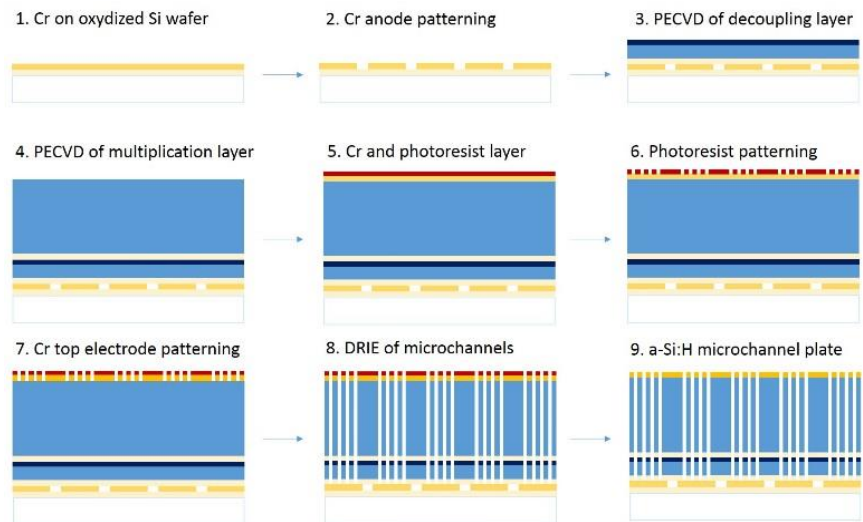

*Figure 2: Schematic representation of the fabrication process of AMCPs. The substrate is an oxidized 6″ silicon wafer, represented by a $SiO_x$ (beige) layer on top of the Si substrate. The Cr anode (dark yellow) is evaporated onto the $SiO_x$. layer. In step 1 → 2, the Cr layer is patterned via lithography. In the following PECVD steps 2 → 3 and 3 → 4, a $SiO_x$ adhesion layer and an a-Si:H (blue) decoupling layer are deposited , followed by another $SiO_x$ adhesion layer and the a-Si:H multiplication layer. Subsequently, the Cr electrode is evaporated on top and coated with a photoresist layer, step 4 → 5. The photoresist and then the Cr layer are patterned in steps 5 → 7. Using Cr as a hard mask, micro channels*

*are micro-machined through the whole layer stack down to the electrodes, step 7 → 8. Remaining resist is removed in step 8 → 9 and the final product is an AMCP.*

A Cr layer of about 80 nm thickness is then evaporated onto the multiplication layer. The metallic layer is used as a hard mask for etching of the micro channels and later acts as an electrode in the AMCP. With a lithographic process, as described for step 2, a 1.2 µm layer of photoresist is coated on top in step 5. The resist is patterned with circular shapes for AMCPs with different diameters ranging from 2 µm to 6 µm in step 6 and after a hard bake, the Cr mask is etched with dry etching in step 7. From here, micro channels are etched into the PECVD layers using state-of-the-art dry reactive ion etching, step 8. Finally, the resist is stripped with acetone from the AMCP in step 9.

Following this fabrication sequence, straight channels with diameters of down to 2.6 µm have been fabricated. SEM images of channels with different diameters in a 55 µm thick AMCP layer stack on the same wafer can be seen in Figure 3. Etching of different channel diameters on the same wafer can be delicate because of lower etch rates for smaller open areas or narrow channels. With the current etching process, channels with 5 µm diameters were etched perfectly straight, see left image of Figure 3. At the same time, the DRIE process still needs to be optimized for smaller channel diameters as shown in the picture on the right. With the current diameter of channels ranging between 2.6 µm and 3.1 µm, we obtained aspect ratio values between 18 and 21. With a thicker a-Si:H layer (100 µm thick AMCPs have been realized on 4″ wafers) we could reach aspect ratio values of up to 40.

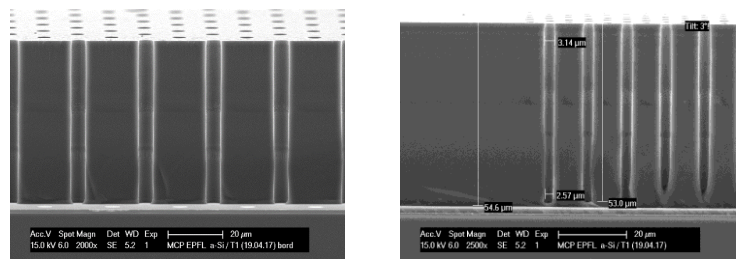

*Figure 3 : SEM images of channels etched for AMCPs. Both structures have been fabricated on the same wafer, during the same process. For the structure in the left image, the realized channel diameter is about 5 µm. What is remarkable here are the straight channels, with almost constant diameter, which reach all the way to the bottom anode. On the right side, an AMCP structure with the present minimal diameter of about 2.6 µm is shown. Here, the micro channels are less homogenous and do not yet reach the bottom electrode. The etching process thus needs to be optimized for small diameters.*

Compared to glass-based MCPs, AMCPs bring the advantage of a rough surface inside the channel, see left image of Figure 4. DRIE walls are known to exhibit a scalloping structure due to the etching process. Using the newer processes, which is able to fabricate straight channel walls, we see that the structure inside the channel is much finer. As secondary emission increases with the active surface area and depends on the nano-curvature of the surface [6], this structure of the walls could have a beneficial effect on the secondary emission strength in AMCPs.

With the photolithographic process used here, the channel diameter of the etching mask could even be reduced from what is currently 2 µm down to 1.5 µm, as see Figure 4 right. For even smaller circular shapes with diameters in the range of hundreds of nanometers, a stepper or e-beam lithography could be used. However, these very narrow channels could be very difficult to etch uniformly on a 6″ wafer.

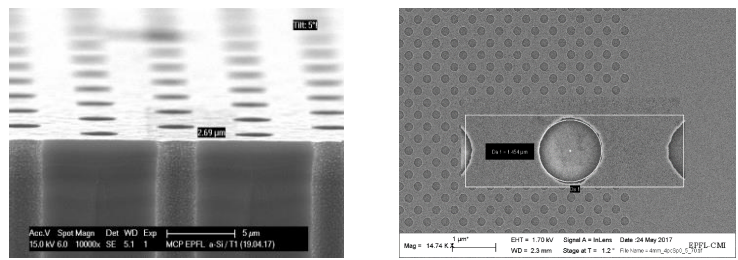

*Figure 4 : Left: SEM image of an AMCP showing the rough surface inside the micro channels. Right: SEM image of a Cr mask produced with the lithographic tools used in this fabrication process. Circular shapes with diameters of down to 1.5 µm are able to be fabricated with the lithographic mask*

4. AMCP gain modeling for high aspect ratio structures

The multiplication gain $G$ of microchannel plates can be expressed depending on their channel aspect ratio and the secondary emission coefficient through the following formula [7]:

$$G = \delta_1 \delta^{n-1} = \delta_1 \left(\frac{V}{n \cdot V_c}\right)^{k(n-1)},$$

where $\delta$ is the secondary emission coefficient, $n$ the number of collisions in one channel and $\delta_1$ is the secondary coefficient of the first collision. For a given material, $k$ is a constant. Also the first crossover potential $V_C$, the minimum potential for secondary emission, is constant within one channel. From this formula, we see that the logarithm of the gain depends linearly on the logarithm of the bias voltage for a given aspect ratio:

$$\log G = k(n-1) \log V - k(n-1) \log n \cdot V_c + \log \delta_1$$

So far, the multiplication gain of AMCPs with aspect ratios of up to 12.5 has been characterized for a primary electron energy of 500 eV [5]. A plot of the logarithm of the gain versus the logarithm of the bias voltage is shown in Figure 5a. From this, we can extract the number of collisions $n$. Figure 4b shows the relation between AMCP aspect ratio and $n$. As for MCPs, $n$ grows linearly. $n$ depends on the aspect ratio $AR$ according to the equation: $n = (0.75 \pm 0.18) \cdot AR - 2.47$. Producing AMCPs with an aspect ratio of 21 would mean a number of collisions of 13, which is starting to be comparable to that of glass-tube MCPs [8], see Figure 5b. The achievable aspect ratio being 40, we can expect a number of collisions of 28.

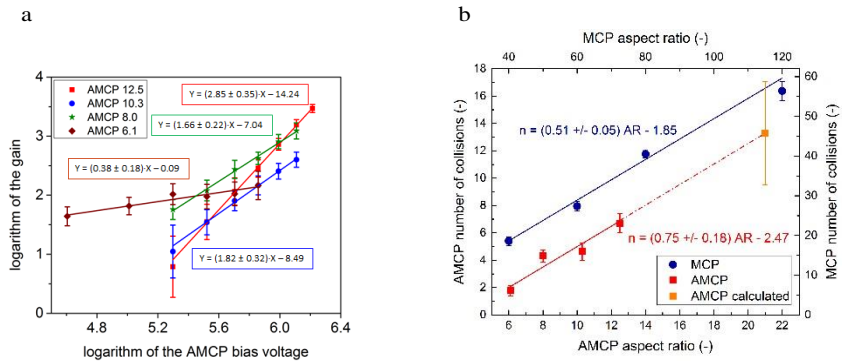

*Figure 5 : a) Measured logarithm of the gain for AMCPs with different aspect ratios. The logarithm of the gain increased linearly with the logarithm of the bias voltage. For each aspect ratio, the number of collisions can be extrapolated. From [5] b) Number of collisions in AMCPs/MCPs plotted versus their respective aspect ratio. As for MCPs, the number of collisions in AMCPs grows linearly with their aspect ratio.*

With the calculated number of collisions $n=13$ and the secondary emission coefficient of $\delta=1.7$ for a-Si:H, a gain value of $G = \delta^n = 10^3$ can be achieved. This shows that AMCPs are on their way to achieve gain values comparable to those of conventional MCPs.

Like borosilicate MCPs, AMCPs can be coated via atomic layer deposition (ALD) with layers of high secondary emissive materials. An optimized ALD coating of 12.5 nm $Al_2O_3$ exhibits a secondary emission yield of $\delta=3.6$ at an incoming electron energy of 400 eV [9]. This could increase the gain tremendously to values of up to $10^7$. This effect is even more pronounced with a higher secondary emissive coefficient of the ALD coating. For 20 nm of ALD-deposited MgO with one cycle of $TiO_2$ and after $Ar^+$ sputtering, a value of $\delta=5.5$ has been measured at a primary electron energy of 500 eV [10]. Such values of secondary emission would allow for very high gain even with moderate AMCP aspect ratios.

5. Conclusion

With a new and improved DRIE process, AMCPs have the potential of achieving gain values comparable to or higher than those of conventional MCPs. The effect of amorphous silicon as a bulk material of MCPs should now be studied. Of particular interest is the transient behavior of AMCPs, because charge replenishment in the channel walls depends on the bulk material. In applications for which a high particle flux, with a high temporal and spatial resolution is needed, detector dead time should be as low as possible. A conventional MCP channel exhibits a dead time in the range of ms. If this time can be improved, current limitations of the data acquisition rate in highly spatially resolved detection can be pushed further.

Not only with additional characteristics but also as an alternative with a considerably easier manufacturing method suitable for large areas, AMCPs could play an important role in modern detector development. As soon as a minimum gain in the range of $10^3$-$10^4$ is overcome, a wide range of applications can be considered. With the novel DRIE process presented here but also with effective ALD coatings the way has been paved in order to fabricate high gain AMCPs that should overcome this minimum.

6. References

[1]     J. Ladislas Wiza, Microchannel plate detectors, Nucl. Instruments Methods. 162 (1979) 587–601. doi:10.1016/0029-554X(79)90734-1.
[2]     O.H.W. Siegmund, J.B. McPhate, S.R. Jelinsky, J. V. Vallerga, A.S. Tremsin, R. Hemphill, H.J. Frisch, R.G. Wagner, J. Elam, A. Mane, Large area microchannel plate imaging event counting detectors with sub-nanosecond timing, IEEE Trans. Nucl. Sci. 60 (2013) 923–931. doi:10.1109/TNS.2013.2252364.
[3]     A. Franco, Y. Riesen, N. Wyrsch, S. Dunand, F. Powolny, P. Jarron, C. Ballif, Amorphous silicon-based microchannel plates, Nucl. Instruments Methods Phys. Res. Sect. A Accel. Spectrometers, Detect. Assoc. Equip. 695 (2012) 74–77.

doi:10.1016/j.nima.2011.11.089.
[4] B. Bläsi, N. Tucher, O. Höhn, V. Kübler, T. Kroyer, C. Wellens, H. Hauser, Large area patterning using interference and nanoimprint lithography, Spie. 9888 (2016) 98880H. doi:10.1117/12.2228458.
[5] A. Franco, J. Geissbühler, N. Wyrsch, C. Ballif, Fabrication and characterization of monolithically integrated microchannel plates based on amorphous silicon, Sci. Rep. 4 (2014) 1–7. doi:10.1038/srep04597.
[6] J. Kawata, K. Ohya, K. Nishimura, Simulation of secondary electron emission from rough surfaces, J. Nucl. Mater. 220–222 (1995) 997–1000. doi:10.1016/0022-3115(94)00460-9.
[7] E.H. Eberhardt, Gain model for microchannel plates, Appl. Opt. 18 (1979) 1418–1423. doi:10.1364/AO.18.001418.
[8] E.H. Eberhardt, An Operational Model for Microchannel Plate Devices, IEEE Trans. Nucl. Sci. 28 (1981) 712–717. doi:10.1109/TNS.1981.4331267.
[9] H. van der Graaf, H. Akhtar, N. Budko, H.W. Chan, C.W. Hagen, C.C.T. Hansson, G. Nützel, S.D. Pinto, V. Prodanović, B. Raftari, P.M. Sarro, J. Sinsheimer, J. Smedley, S. Tao, A.M.M.G. Theulings, K. Vuik, The Tynode: A new vacuum electron multiplier, Nucl. Instruments Methods Phys. Res. Sect. A Accel. Spectrometers, Detect. Assoc. Equip. 847 (2017) 148–161. doi:10.1016/j.nima.2016.11.064.
[10] S.J. Jokela, I. V. Veryovkin, A. V. Zinovev, J.W. Elam, A.U. Mane, Q. Peng, Z. Insepov, Secondary Electron Yield of Emissive Materials for Large-Area Micro-Channel Plate Detectors: Surface Composition and Film Thickness Dependencies, Phys. Procedia. 37 (2012) 740–747. doi:10.1016/j.phpro.2012.03.718.

Acknowledgments

We are thankful to our colleagues, in particular P.-A. Clerc, from the Swiss Center for Electronics and Microtechnology for performing the dry reactive ion etching, their expertise on lithography processes and for providing SEM images of all DRIE results shown in this publication. We thank the photolithography team of the Center of Micronanotechnology of the Ecole Polytechnique Fédérale de Lausanne for their service and for providing the SEM image of their lithography mask resolution. Results presented here were obtained with the financial support of the Swiss National Science Foundation project 200021_162525/1.